\documentclass[lettersize,journal]{IEEEtran}

\usepackage[utf8]{inputenc}
\usepackage[final, pdftex]{graphicx}
\graphicspath{ {./epsfigs/} }
\usepackage[reqno]{amsmath}
\usepackage{amssymb}
\usepackage{amsthm}
\usepackage{subfig}
\usepackage{epstopdf}
\usepackage{setspace}
\usepackage[T1]{fontenc}
\usepackage{color,soul}
\usepackage{multirow}
\usepackage{array}
\usepackage{cite}
\usepackage{svg}
\usepackage{verbatim}
\usepackage{enumitem,kantlipsum}
\usepackage{blindtext}
\usepackage{makecell}
\usepackage{anyfontsize}

\usepackage{geometry}
\usepackage{xcolor}

\usepackage[ruled]{algorithm2e}
\usepackage[hyphens]{url}

\newcolumntype{C}[1]{>{\centering\arraybackslash}p{#1}}

\newcommand{\ignore}[1]{}

\begin{document}

\title{Deep OFDM Channel Estimation: Capturing Frequency Recurrence}

\author{\large{Abu Shafin Mohammad Mahdee Jameel$^{\dagger}$, {\em Student Member, IEEE}}, \large{Akshay Malhotra$^{\ddagger}$, {\em Senior Member, IEEE}}, \large{Aly El Gamal$^{\dagger}$, {\em Senior Member, IEEE}}, and \large{Shahab Hamidi-Rad$^{\ddagger}$, {\em Member, IEEE} 
\vspace{-5mm}}
\thanks{$^{\dagger}$ The authors are with the School of Electrical and Computer Engineering, Purdue University. West Lafayette, Indiana, USA (e-mail: \{amahdeej, elgamala\}@purdue.edu)

$^{\ddagger}$ The authors are with InterDigital Communications, Inc. Los Altos, CA, USA(e-mail: \{akshay.malhotra, shahab.hamidi-rad\}@interdigital.com).}}

\maketitle

\begin{abstract}
In this paper, we propose a deep-learning-based channel estimation scheme in an orthogonal frequency division multiplexing (OFDM) system. Our proposed method, named Single Slot Recurrence Along Frequency Network (SisRafNet), is based on a novel study of recurrent models for exploiting sequential behavior of channels across frequencies. Utilizing the fact that wireless channels have a high degree of correlation across frequencies, we employ recurrent neural network techniques within a single OFDM slot, thus overcoming the latency and memory constraints typically associated with recurrence based methods. The proposed SisRafNet delivers superior estimation performance compared to existing deep-learning-based channel estimation techniques and the performance has been validated on a wide range of 3rd Generation Partnership Project (3GPP) compliant channel scenarios at multiple signal-to-noise ratios.

\end{abstract}


\IEEEpeerreviewmaketitle
\section{Introduction}

\IEEEPARstart{O}{rthogonal} frequency division multiplexing (OFDM) is among the most important building blocks for a modern wireless communication system. Owing to its robustness to frequency selective fading and its bandwidth efficiency, OFDM has been adopted as an integral part of the Long-Term Evolution (LTE), and Fifth Generation New Radio (5G NR) standards. In a wireless system, the transmitted signal is typically subjected to distortion, fading, and path loss on account of the characteristics of the propagation channel. The received signal is also corrupted by additive noise and interference, further degrading the quality of the signal. For reliable recovery of the transmitted signal in an OFDM system, the degradation associated with the propagation channel needs to be effectively estimated and compensated for. 

Channel estimation in OFDM settings is well-studied and several approaches have been proposed towards solving this problem---employing statistical estimators \cite{zhou2003first},  matrix factorization \cite{Suga2019factorization}, low-rank matrix completion \cite{li2017millimeter}, and other model-based techniques. One of the most well-known approaches for channel estimation is the minimum mean square error (MMSE) estimator, wherein the second-order statistics of the channel and the noise are utilized in the estimation process. On the other hand, when the channel statistics are unavailable, a commonly used approach is the least squares (LS) estimator. MMSE based approaches generally exhibit better performance compared to LS estimators, but with slightly higher complexity. 

Recently, deep-learning-based techniques have been proposed as an alternative to traditional approaches. Initial approaches focused on reusing ideas from the image processing field as the channel estimation problem is conceptually similar to the upsampling problem usually referred to as image super-resolution in literature \cite{soltani2019deep, shen2023deep,   zhang2020lsrn}. These approaches predominantly utilize convolutional neural networks (CNN) due to their ability to learn and uncover local patterns. In ChannelNet \cite{soltani2019deep} the authors proposed a two-stage network which is similar to an image super-resolution network. At the first stage, the channel estimates at pilot locations are upsampled to the complete channel estimate and at the second stage, a denoising network is used to filter out the noise.  In \cite{shen2023deep}, an end-to-end trained network is used to solve the channel estimation problem in a setup equipped with reconfigurable intelligent surfaces. CNN based methods have also been shown to perform well when imperfect channel information is available \cite{ge2021deep}. The use of CNN based autoencoders \cite{lin2020novel} and generative adversarial networks (GANs) \cite{balevi2021wideband} have also been explored for the channel estimation problem.

Although CNN based methods have been utilized for channel estimation with some success, they are not particularly well suited for modeling the long term variations across time and frequency in the channel matrix, especially when the user equipment (UE) is mobile. Hence, there has been recent focus on utilizing long term temporal correlations using recurrent neural network (RNN) blocks \cite{liao2019chanestnet}. In \cite{jiang2021dual}, separate CNN networks are employed on the Spatial-Frequency (SF) and Angular-Domain (AD) signals for channel estimation. The authors further utilize RNN networks to capture time domain correlations. In \cite{yang2021deep}, separate CNN based systems are employed which focus on 1D slices from either the time or frequency axis, and their results are combined. However, this implementation does not explore the effect of recurrence along time or frequency axis. Utilizing time domain recurrence, long short-term memory (LSTM) based systems are shown to outperform purely CNN based systems \cite{ farsad2018neural}. Recently, gated recurrent unit (GRU) based methods, which use computationally less complex RNN blocks, have been shown to outperform LSTM based systems \cite{hou2022gru} while offering faster performance.

While these works utilize RNN based methods across multiple OFDM slots for improved estimation, they continue to rely on CNN-based filters, that are capable of capturing only localized correlations, for feature extraction across frequencies or sub-carriers.  
Additionally, deploying such multiple-slot RNN-based solutions requires storing the estimates from past frames in memory, which imposes additional memory and computational complexity on generally resource-constrained user equipment (UE).  

In this paper, we propose a channel estimation method based on a novel investigation of recurrent models for capitalizing on the sequential nature of the channel across frequencies (or sub-carriers). 

The proposed channel estimation framework outperforms the previous approaches both in noise-free and noisy environments. The contributions of this work are listed below:
\begin{itemize}
    \item Our novel framework aims to leverage neural recurrence to capture correlations across frequencies (sub-carriers) to improve the channel estimation performance in OFDM systems. Unlike past applications of recurrent neural networks for applying recurrence across time, our work explores the usage of these models along the frequency axis.

    \item We propose a single slot approach with frequency domain RNNs (GRU) that provides state-of-the-art results across a wide range of SNRs compared to existing methods in the literature. Since the RNN is utilized along the frequency (or sub-carrier) axis, there is no associated latency with this method and it operates standalone on a single slot. 

    \item We show the performance of the proposed SisRafNet in diverse settings consisting of multiple 3GPP-compliant channel models. The proposed scheme is compared against popular deep learning-based solutions such as ChannelNet \cite{soltani2019deep}, ChanEstNet \cite{liao2019chanestnet} and SRDnNet \cite{shen2023deep}.

\end{itemize}

\section{System Model and Problem Formulation}

Consider a downlink orthogonal frequency division multiplexing (OFDM) communication system with $N_t$ transmit antennas and $N_r$ receive antennas. The OFDM transmission is characterized by $K$ sub-carriers and  $N_s$ OFDM symbols, thus the transmission at each OFDM slot from each transmit antenna is represented as a matrix of dimensions $K \times N_s$. 

To simplify the discussion and without loss of generality, we consider a setting with a single transmit antenna and a single receive antenna $N_t = N_r =1$. The overall transmission at a given OFDM slot can be expressed as:
\begin{equation}
    \mathbf{Y}_{[i,k]} = \mathbf{H}_{[i,k]}\mathbf{X}_{[i,k]}+\mathbf{Z}_{[i,k]},
\end{equation}
where \(\mathbf{Y}_{[i,k]}\) and \(\mathbf{X}_{[i,k]}\), are the received and transmitted OFDM symbols, respectively.  \(\mathbf{H}_{[i,k]}\) is the $(i, k)$ element of channel matrix $\mathbf{H} \in \mathbb{C} ^{K \times N_s}$ and represents the frequency-time response of the wireless communication channel. The additive gaussian noise at the receiver is given  as \(\mathbf{Z}_{[i,k]}\) and is assumed to be independent and identically distributed (i.i.d) with zero mean and variance $\sigma^2$.

In order to obtain the transmitted signal using the received signal, the channel response information associated with each slot needs to be estimated for all sub-carriers and for all OFDM symbols; thus, assuming knowledge of \(\mathbf{Y}_{[i,k]}\) and \(\mathbf{X}_{[i,k]}\), we can calculate \(\mathbf{H}_{[i,k]}\) for a noise-free channel. 

To determine the channel, a common practice is to transmit \textit{pilot signals} known apriori at the receiver as $\mathbf{X}_{{[i,k]}}$  and utilize them to find a least squares (LS) estimate of the channel. 

\begin{figure}[t]
    \centering
    \includegraphics[width=0.30\textwidth]{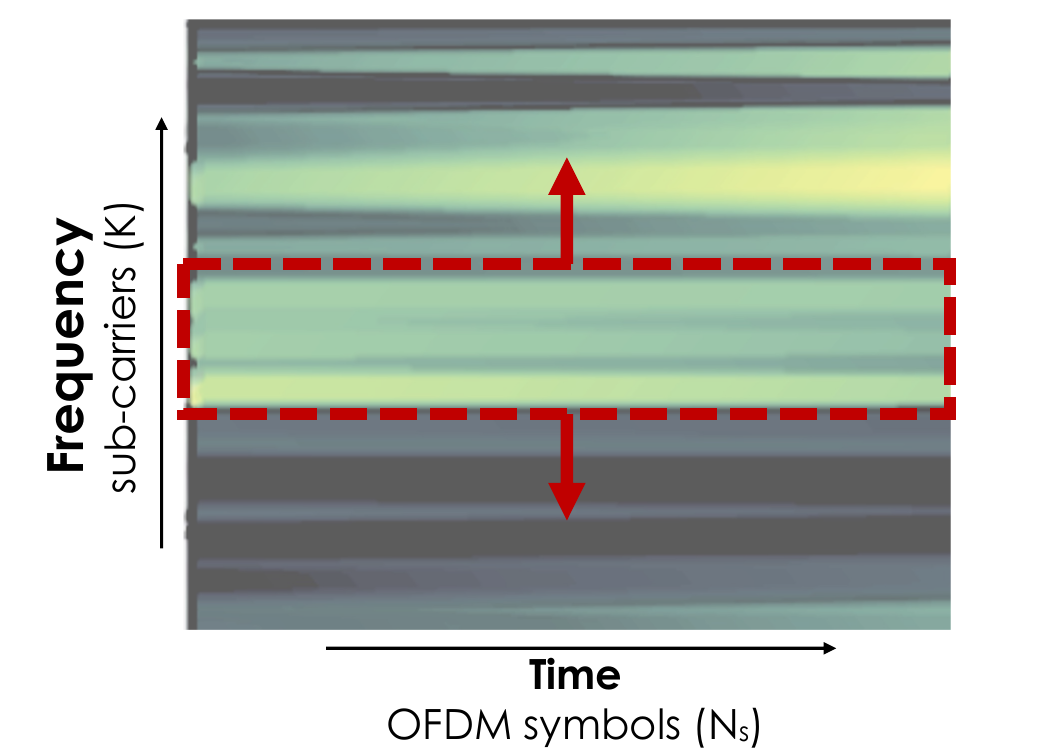}
    \caption{Direction of recurrent layer operations for proposed SisRafNet with recurrence across frequency.\vspace{-10pt}}
    \label{fig:recurrence_dir}
\end{figure}

\begin{equation}
    \Tilde{\textbf{H}}_{[i,k]}^{LS} = \arg \min _{\textbf{H}_{[i,k]}} | \textbf{Y}_{{[i,k]}} - \textbf{H}_{[i,k]}\textbf{X}_{{[i,k]}} |^2.
\end{equation}
Since sending pilot signals for a high number of entries in $K \times N_s$ data grid would strongly restrict the throughput of the communication link, only few pre-determined entries of the grid contain pilot signals, whereas the others contain the data to be transmitted. Let $\mathcal{P}$ be the set containing the $[i,k]$ locations of the entries where pilot signals are transmitted. At the receiver, LS based channel estimation is performed to estimate $\Tilde{\mathbf{H}}_p \forall ~ p \in \mathcal{P}$ and interpolation is employed to estimate the channel corresponding to each entry of ${\mathbf{H}}$ \cite{soltani2019deep}.

\begin{figure*}[h]
    \centering
    \includegraphics[width=0.98\textwidth,height =1.3in]{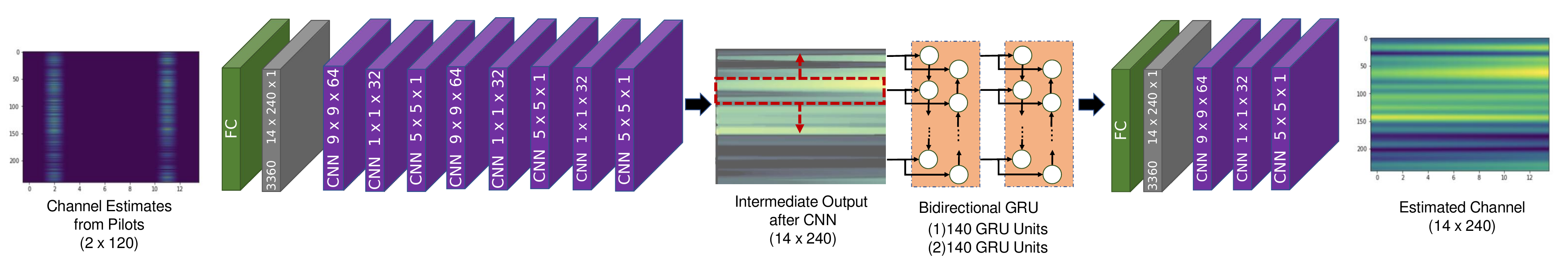}
    \caption{Architecture of the Proposed SisRafNet with Bidirectional GRU layers.}
    \label{fig:network}
\end{figure*}

In this work, we are interested in developing a deep learning based model $f(\cdot)$ to utilize the least squares based channel estimates at pilot locations, $\Tilde{\mathbf{H}}_p = \hat{K} \times \hat{N}_s \subset K \times N_s$,  and estimate the entire $K \times N_s$ channel matrix ${\mathbf{H}}$, such that $f(\Tilde{\mathbf{H}}_p ~|~ p \in \mathcal{P} ) \rightarrow \ \mathbf{H}$. Since the channel response is sequential across sub-carriers, we aim to utilize recurrent neural networks to effectively model the long and short-term correlation in the data for channel estimation.  

We estimate the real and imaginary parts of the complex channel matrix separately. In line with the formulation in \cite{soltani2019deep}, we train a single deep learning model $f(\cdot)$, that independently utilizes the 
real or imaginary component of the pilots $\Tilde{\mathbf{H}}_p$ to estimate the 
real or imaginary component of the channel matrix $\mathbf{H}$. 

\section{Proposed Solution}

In our approach to the problem of channel estimation using noisy estimates at the pilot locations, we propose to utilize the inherent short and long term correlations across frequency (sub-carriers) within an OFDM slot.

\subsection{Frequency Domain Recurrence}
While operating on a single OFDM slot, the dimensionality of the $ K \times N_s$ channel matrix is heavily skewed, $K > N_s$. As in the 5G case, the OFDM slot length is typically $N_s = 14$, whereas the number of sub-carriers can be much larger ($K=240$ sub-carriers for the 5G setup investigated in this paper).Therefore, given the small value of $N_s$, CNNs are sufficiently capable of capturing the local relationships within the slot. On the other hand, since the dimensionality of $K$ can be significantly large and since long-term correlations may exist within non-neighboring sub-carriers, CNN may perform poorly. Consequently, we propose to employ recurrent neural networks along the sub-carriers or frequencies (i.e.,  along the Y-axis in Fig. \ref{fig:recurrence_dir}), in conjunction with convolutional networks that uncover local trends. This utilization of recurrence across frequencies provides large gains in the channel estimation performance.

\subsection{Proposed Single OFDM Slot Model}

In this model, we utilize the proposed implementation of recurrent neural networks across the frequency blocks, as shown in Fig. \ref{fig:network}. On the left, we illustrate the positions of the pilot elements (and their magnitude) with respect to the full OFDM slot. The input to the system contains only a $\hat{K} \times \hat{N_s}$ ($\hat{K} = 120$ sub-carriers and  $\hat{N_s} = 2$ OFDM symbols) least squares estimate of the channel, obtained using the pilot signals. This estimate is noisy as it is affected by the noise captured at the receiver. In order to estimate the full $240 \times 14$ channel, we utilize a deep learning model with fully connected, convolutional, and recurrent layers. The recurrent neural networks deployed here use bidirectional GRUs, which are computationally lighter compared to the popular LSTM recurrent units, while providing similar levels of performance \cite{bai2019deep}. The channel estimation network consists of 8 CNN layers, followed by 2 GRU layers (that operate along the frequency domain), followed by a fully connected layer and 3 CNN layers. 

Fig. \ref{fig:network} showcases the overall architecture with an example input and output. The intermediate output of the CNN layers is also showcased to better explain the frequency domain operation of the GRU layers. At the GRU layers, the processing is sequential and is signified by the red sliding window. The contents of the sliding window are sequentially processed by the GRU layers, and at each instant, the window slides down across frequency blocks (or up, depending on the direction of processing). Due to the bidirectional nature of the GRU block, there is a forward GRU and a backward GRU calculation, where in one case the GRU memory scans from lower to higher frequencies while in the other case the memory scans from higher to lower frequencies. The outputs of both GRU directions are combined and passed through a fully connected layer, followed next by 3 CNN layers to arrive at the final output.

\section{Results}

\subsection{Dataset}
We present our findings in a simulation setting with CDL-A and CDL-D cluster delay line channel models as described in 3GPP. We use two different delay spreads (30 ns and 300 ns) and two different Doppler shift velocities (3 km/hr and 30 km/hr). This gives us a to total of 8 different channel settings. A single transmit and single receive antenna setting is considered. A carrier frequency of 3.5 GHz with 240 sub-carriers and a sub-carrier spacing of 30 KHz is selected. For training and validation, we consider a total of 800 channel realizations (100 from each individual setting), where each channel realization consists of sequential channel data across 100 OFDM slots (each OFDM slot has 14 OFDM symbols and 240 sub-carriers), thus a total of 80,000 OFDM slots of channel data has been considered. For all these cases, the dataset contains both the noisy channel estimate at pilot points and the ground truth of the channel. We have used 80\% of the realizations for training, 10\% for validation, and 10\% for testing. As the mean square error (MSE) is widely accepted as the measure of channel estimation performance, the MSE loss function along with the Adam optimizer are employed during neural network training.

\subsection{Simulation Results}

To show the efficacy of the proposed approach we compare our proposed SisRafNet with ChannelNet \cite{soltani2019deep}, which is a popular CNN based channel estimation method; SRDnNet \cite{shen2023deep}, which is a recent end-to-end method inspired by Channelnet and ChanEstNet \cite{liao2019chanestnet}, where in addition to CNN, bidirectional LSTM layers are also utilized in the deep learning model to better utilize the recurrence across time. For ChannelNet, we used the implementation shared by the authors in their GitHub page. For ChanEstNet and SRDnNet, we implemented the model described in the paper, tuned to match our input data.

\begin{figure}[t]
    \centering
    \includegraphics[width=0.480\textwidth]{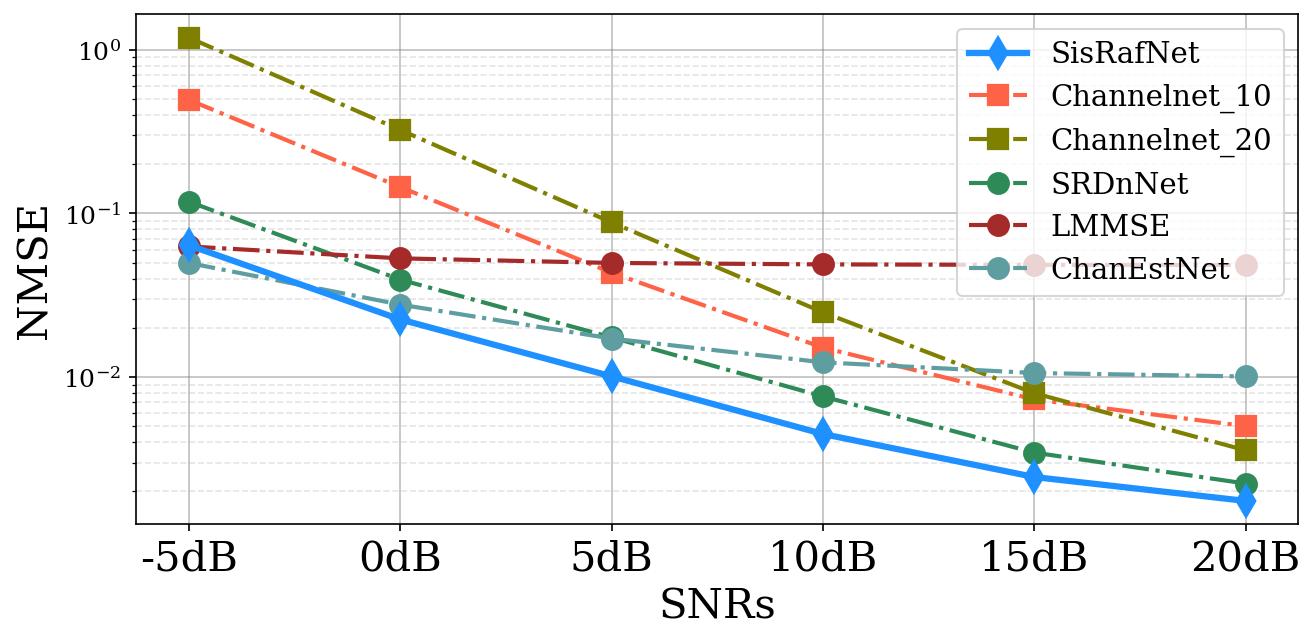}
    \caption{NMSE performance comparison of SisRafNet.}
    \label{fig:results}
\end{figure}

In Fig. \ref{fig:results}, we showcase the performance of the proposed SisRafNet channel estimation scheme. Here, the normalized mean square error (NMSE) between an estimated channel matrix and the ground truth is presented for SisRafNet as well as benchmark algorithms. The plot shows the average NMSE across the test set. For the ChannelNet algorithm, there are two networks as proposed in \cite{soltani2019deep}, one trained using training samples estimated at 10 dB SNR, and the other at 20 dB SNR. The authors proposed manually switching between these two networks based on the estimated SNR. On the other hand, none of the other methods require an SNR estimate. The ChanEstNet algorithm is a multi OFDM slot algorithm, while ChannelNet and the proposed SisRafNet both require only a single OFDM slot. We also include results from a traditional linear minimum mean square error (LMMSE) estimator based on \cite{cho2010mimo,savaux2017lmmse}.

\begin{itemize}
    \item From Fig. \ref{fig:results}, it can be seen that the proposed SisRafNet method outperforms traditional LMSSE, SRDnNet and both the ChannelNet models across all SNR ranges.
    
    \item SisRafNet outperforms ChanEstNet by a wide margin at high SNR levels. However, at low SNRs of under 0 dB, the performance of the 2 schemes are very close. It is important to note that ChanEstNet utilizes pilots from 10 past OFDM slots for the channel estimation, while SisRafNet needs only a single OFDM slot.

\end{itemize}

\subsection{SNR Robust Training}

\begin{figure}[t]
    \centering
    \includegraphics[width=0.480\textwidth]{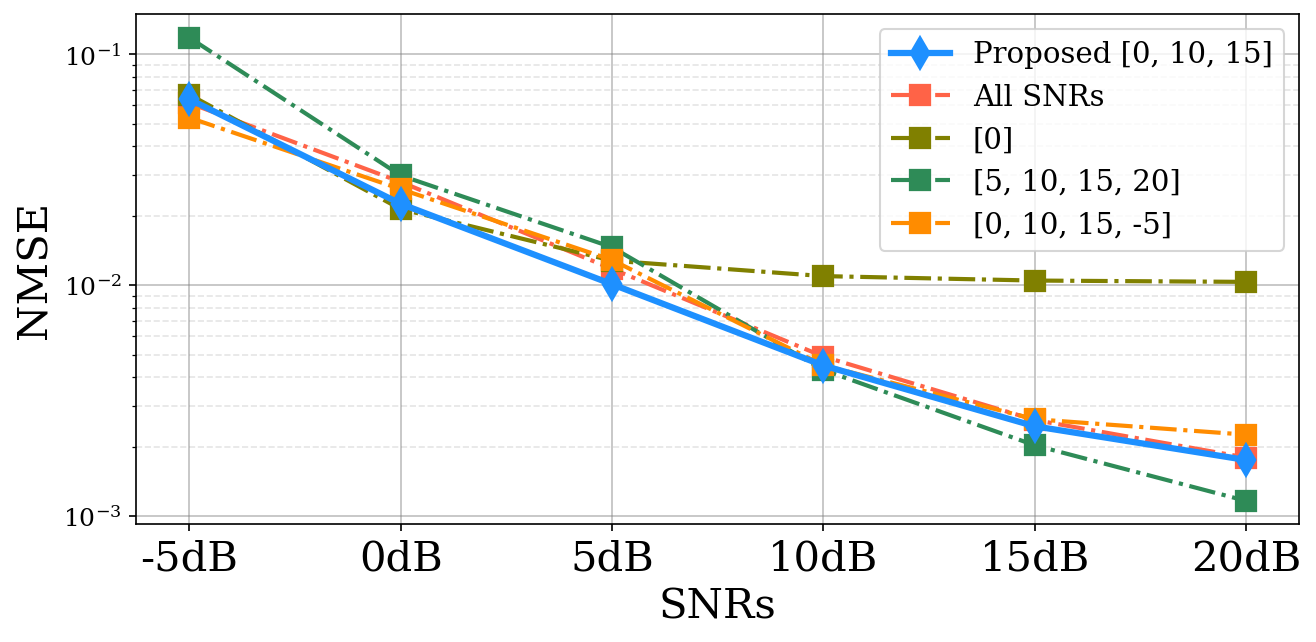}
    \caption{Performance when training with samples from different SNR combinations.}
    \label{fig:snrboostresults}
\end{figure}

In Fig. \ref{fig:snrboostresults}, we investigate the effect of training with samples from different SNR conditions. In all these cases, the test dataset is the same standard one used before, with representation from all SNR ranges. Firstly, we show results where we take equal number of training samples from all SNR levels (-5 dB to 20 dB in 5 dB increments). With the assumption that training samples under severe noise might not be useful due to corruption, we also test the case where we train with only samples with SNR greater than 5 dB, and while it provides superior performance under low noise (SNR > 10 dB), the performance deteriorates in noisy environments (SNR < 5 dB). On the other hand, training with only a single noisy SNR (0 dB) works for similarly noisy environments, but fails in environments with low noise. Finally, we also show results for two good performing combinations suggested by the SNR Boost algorithm, adapted from \cite{wang2020efficient}. In the SNR Boost algorithm, a greedy search is performed to find out the best possible combination of training SNRs. For a set of 3 best SNRs, the best results are obtained for a mixture of $\{0,10,15\}$ dB samples. We can see that a set of 4 SNRs (a mixture of $\{0,10,15,-5\}$ dB samples) does not improve performance. From the results in Fig. \ref{fig:snrboostresults}, we can see that it is possible to achieve similar result characteristics when training using a subset of SNR ranges. This observation is quite useful in two fronts. Firstly, it hints that the network, when trained with carefully selected SNR ranges, can still perform well in other SNR ranges. Secondly, it decreases the complexity involved in obtaining training samples. As a result, the proposed models in this paper are trained with only the mixture of $\{0,10,15\}$ dB samples.

\subsection{Generalization}

\begin{figure}[t]
    \centering
    \includegraphics[width=0.480\textwidth]{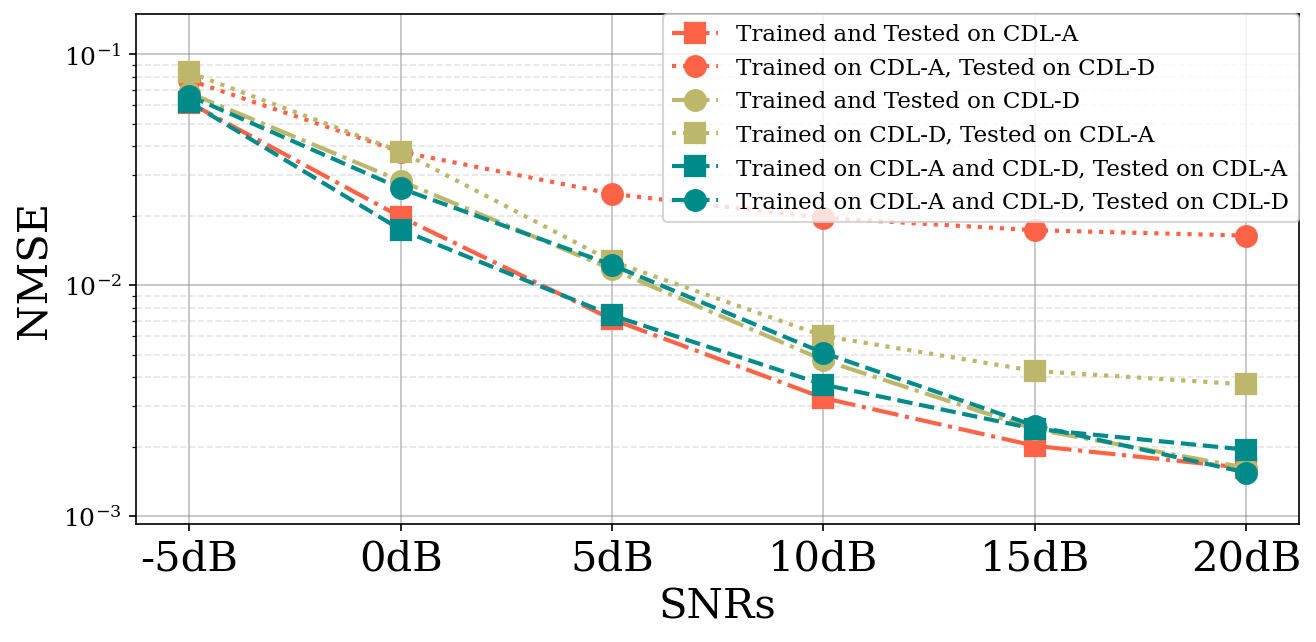}
    \caption{Generalization across different channel models.}
    \label{fig:datasetresults}
\end{figure}

\begin{figure}[t]
    \centering
    \includegraphics[width=0.480\textwidth]{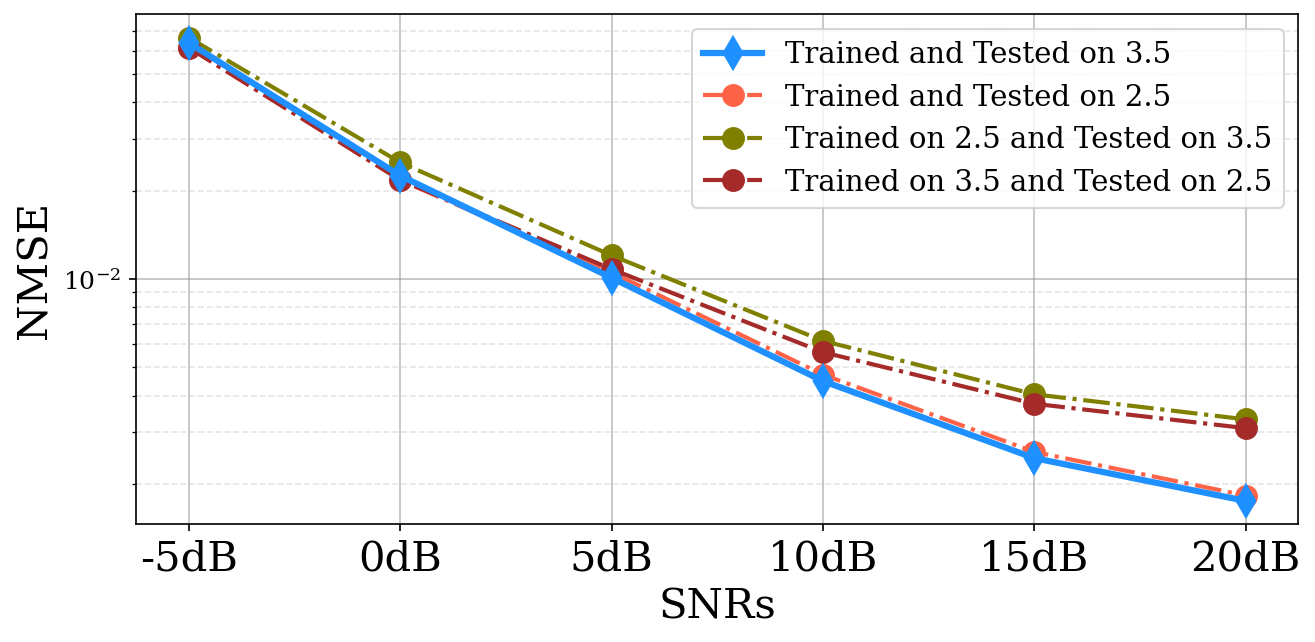}
    \caption{Generalization across different center frequencies.}
    \label{fig:centfreq}
\end{figure}

First, we explore the generalization capability of the proposed SisRafNet across the two different channel models considered in this work: (i) CDL-A, which is a non-line-of-sight (NLOS) channel model, and (ii) CDL-D which is a multipath channel model with both LOS and NLOS paths. For both cases, we consider a combination of different delay spreads (30 ns and 300 ns) and doppler shift velocities (3 km/hr and 30 km/hr).

As seen from Fig. \ref{fig:datasetresults}, the models trained and tested in the within-distribution case (training and test data coming from the same channel model) perform well for both CDL-A and CDL-D scenarios. Although, the performance in the CDL-A dataset is a bit better than CDL-D. Additionally, the model trained with CDL-D does a better job in the out-of-distribution case when tested on CDL-A data. The model trained with the CDL-A dataset has poor generalization performance when tested on CDL-D data, pointing to the fact that the CDL-D is more diverse and better for training  models with better generalization performance. The model trained on a combined dataset mimics the in-distribution performances of models trained on only one dataset.

Next, in Fig. \ref{fig:centfreq}, we investigate the performance of SisRafNet for data sets with different center frequencies. The change in center frequency has almost no effect on the performance of the model when inference is performed within-distribution. In the out-of-distribution case, the performance remains unaffected at the low SNRs. At the higher SNRs a small drop in performance is observed.

\subsection{Effect of Pilot Configurations}

\begin{figure}[t]
    \centering
    \includegraphics[width=0.480\textwidth]{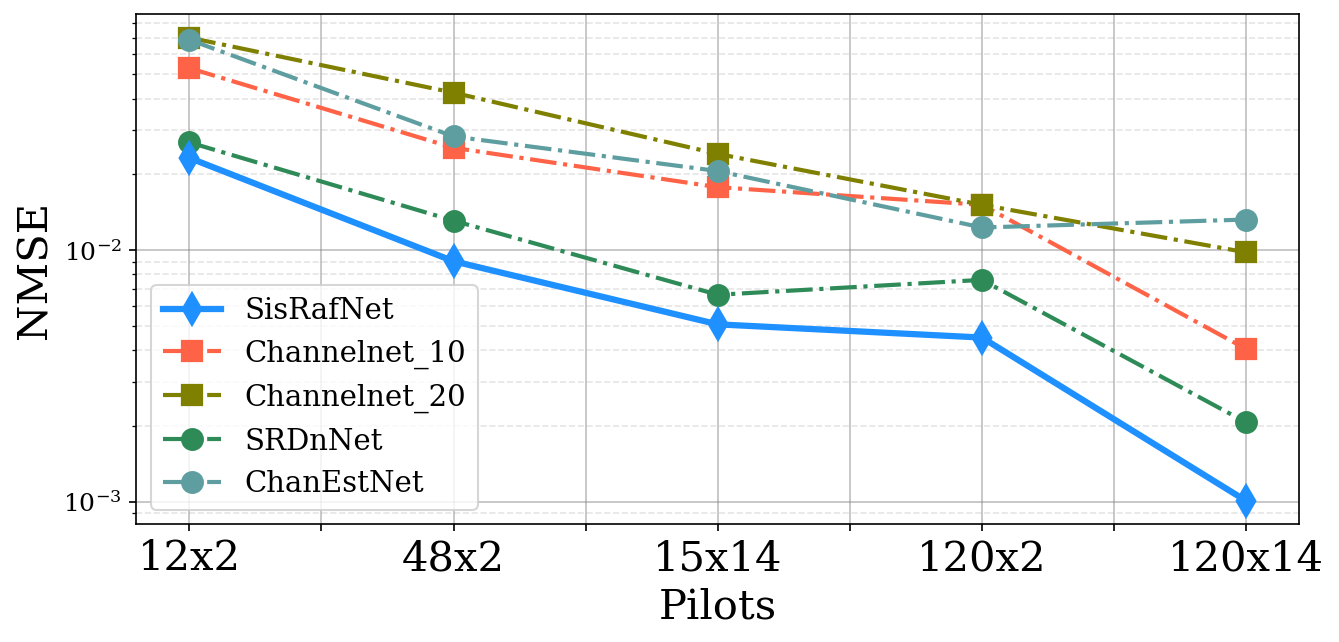}
    \caption{Performance comparison of different deep learning models with different pilot configurations in 10dB SNR.}
    \label{fig:pilots_10dB}
\end{figure}

In Fig. \ref{fig:pilots_10dB}, we present the performance of SisRafNet and the benchmark models for five pilot configurations under 10dB noise. We can see that the performance of the different models show a similar trend of change across different pilot configurations, and SisRafNet is the best performing model in all pilot configurations. In our experiments, the proposed method outperformed the other methods across a diverse configuration of SNR ranges and pilot configurations.

\subsection{Performance Characteristics}

In Table \ref{table:performance}, we present the inference time, Million Floating Point Operations Per Second (MegaFlops), and memory of past slots required to get a channel estimation for a single OFDM slot. The machine where the tests were performed has an Intel Xeon Gold 6142 CPU with 64 cores, 256GB of RAM, and Nvidia Tesla P100 GPU with 16GB of memory. We can observe that SisRafNet outperforms all other models in terms of inference speed. The CPU-bound interpolation step in channelnet makes it the slowest. ChanEstNEt has the lowest MegaFlops count, but SisRafNet compares favorably against Channelnet and SRDnNet. ChanEstNet still has higher inference time than SisRafNet, and needs a memory of 10 OFDM slots worth of pilots, while all other methods can estimate a channel from only one slot.

\section{Conclusion and Future Work}

In this paper, a noise robust OFDM channel estimation scheme using a combination of convolutional and recurrent neural networks was presented. We developed a neural network model that can utilize the frequency correlation present in a single OFDM slot. The proposed CNN and GRU based SisRafNet network does not suffer from the latency and memory constraints that are inherent in recurrent network applications across time. As a future direction, the proposed solution of frequency domain recurrence can be further leveraged to exploit improved performance in other wireless applications including predicting the future channel given the current channel, channel state information (CSI) compression and beamforming. We also plan to explore the performance of the proposed approach in multi-antenna channel systems and study the generalization performance of deep learning-based solutions in over-the-air data with imperfect ground truth availability.

\noindent
\begin{table}[t]
\caption{Performance Characteristics for Different Models}
\label{table:performance}
    \begin{tabular}{|@{\hskip 0mm}p{16mm}|p{7mm}|p{11mm}|p{11mm}|p{9mm}|p{8mm}|}
    \hline
        & \begin{tabular}[c]{@{}l@{}}SisR- \\ afNet\end{tabular} & \begin{tabular}[c]{@{}l@{}}Chann- \\ elnet\_10\end{tabular}     & \begin{tabular}[c]{@{}l@{}}Chann- \\ elnet\_20\end{tabular}     & \begin{tabular}[c]{@{}l@{}}Chan- \\ EstNet\end{tabular} & \begin{tabular}[c]{@{}l@{}}SRD- \\ nNet\end{tabular}          \\ 
        \hline
        Inf.Tim(ms) & \multicolumn{1}{c|}{2.05} & \multicolumn{1}{c|}{29.55} & \multicolumn{1}{c|}{29.59} & \multicolumn{1}{c|}{4.83} & \multicolumn{1}{c|}{4.36} \\ 
        \hline
        MegaFlops & \multicolumn{1}{c|}{501} & \multicolumn{1}{c|}{5787} & \multicolumn{1}{c|}{5787} & \multicolumn{1}{c|}{10.16} & \multicolumn{1}{c|}{4482}\\ 
        \hline
        Mem(Slots) & \multicolumn{1}{c|}{1} & \multicolumn{1}{c|}{1} & \multicolumn{1}{c|}{1} & \multicolumn{1}{c|}{10} & \multicolumn{1}{c|}{1}\\
        \hline
    \end{tabular}
\end{table}

\bibliographystyle{IEEEtran.bst}
\bibliography{IEEEabrv,ofdm_rnn}

\begin{thebibliography}{10}
\providecommand{\url}[1]{#1}
\csname url@samestyle\endcsname
\providecommand{\newblock}{\relax}
\providecommand{\bibinfo}[2]{#2}
\providecommand{\BIBentrySTDinterwordspacing}{\spaceskip=0pt\relax}
\providecommand{\BIBentryALTinterwordstretchfactor}{4}
\providecommand{\BIBentryALTinterwordspacing}{\spaceskip=\fontdimen2\font plus
\BIBentryALTinterwordstretchfactor\fontdimen3\font minus \fontdimen4\font\relax}
\providecommand{\BIBforeignlanguage}[2]{{%
\expandafter\ifx\csname l@#1\endcsname\relax
\typeout{** WARNING: IEEEtran.bst: No hyphenation pattern has been}%
\typeout{** loaded for the language `#1'. Using the pattern for}%
\typeout{** the default language instead.}%
\else
\language=\csname l@#1\endcsname
\fi
#2}}
\providecommand{\BIBdecl}{\relax}
\BIBdecl

\bibitem{zhou2003first}
G.~T. Zhou, M.~Viberg, and T.~McKelvey, ``A first-order statistical method for channel estimation,'' \emph{{IEEE Signal Processing Letters}}, vol.~10, no.~3, pp. 57--60, 2003.

\bibitem{Suga2019factorization}
N.~Suga, R.~Sasaki, and T.~Furukawa, ``{Channel estimation using matrix factorization based interpolation for OFDM systems},'' in \emph{2019 IEEE 90th Vehicular Technology Conference}, pp. 1--5.

\bibitem{li2017millimeter}
X.~Li, J.~Fang, H.~Li, and P.~Wang, ``Millimeter wave channel estimation via exploiting joint sparse and low-rank structures,'' \emph{IEEE Transactions on Wireless Communications}, vol.~17, no.~2, pp. 1123--1133, 2017.

\bibitem{soltani2019deep}
M.~Soltani, V.~Pourahmadi, A.~Mirzaei, and H.~Sheikhzadeh, ``Deep learning-based channel estimation,'' \emph{IEEE Communications Letters}, vol.~23, no.~4, pp. 652--655, 2019.

\bibitem{shen2023deep}
W.~Shen, Z.~Qin, and A.~Nallanathan, ``Deep learning for super-resolution channel estimation in reconfigurable intelligent surface aided systems,'' \emph{IEEE Transactions on Communications}, vol.~71, no.~3, pp. 1491--1503, 2023.

\bibitem{zhang2020lsrn}
S.~Zhang, Y.~Liu, Q.~Shi, S.~Xu, and S.~Cao, ``{LSRN}: A recurrent residual learning framework for continuous wireless channel estimation using super-resolution concept,'' \emph{IEEE Access}, vol.~8, pp. 38\,098--38\,111, 2020.

\bibitem{ge2021deep}
L.~Ge, Y.~Guo, Y.~Zhang, G.~Chen, J.~Wang, B.~Dai, M.~Li, and T.~Jiang, ``Deep neural network based channel estimation for massive {MIMO-OFDM} systems with imperfect channel state information,'' \emph{IEEE Systems Journal}, vol.~16, no.~3, pp. 4675--4685, 2022.

\bibitem{lin2020novel}
B.~Lin, X.~Wang, W.~Yuan, and N.~Wu, ``A novel {OFDM} autoencoder featuring {CNN}-based channel estimation for internet of vessels,'' \emph{IEEE Internet of Things Journal}, vol.~7, no.~8, pp. 7601--7611, 2020.

\bibitem{balevi2021wideband}
E.~Balevi and J.~G. Andrews, ``Wideband channel estimation with a generative adversarial network,'' \emph{IEEE Transactions on Wireless Communications}, vol.~20, no.~5, pp. 3049--3060, 2021.

\bibitem{liao2019chanestnet}
Y.~Liao, Y.~Hua, X.~Dai, H.~Yao, and X.~Yang, ``{ChanEstNet}: A deep learning based channel estimation for high-speed scenarios,'' in \emph{ICC 2019 - 2019 IEEE International Conference on Communications (ICC)}, 2019, pp. 1--6.

\bibitem{jiang2021dual}
P.~Jiang, C.-K. Wen, S.~Jin, and G.~Y. Li, ``Dual {CNN}-based channel estimation for {MIMO-OFDM} systems,'' \emph{IEEE Transactions on Communications}, vol.~69, no.~9, pp. 5859--5872, 2021.

\bibitem{yang2021deep}
A.~Yang, P.~Sun, T.~Rakesh, B.~Sun, and F.~Qin, ``Deep learning based {OFDM} channel estimation using frequency-time division and attention mechanism,'' in \emph{2021 IEEE Globecom Workshops (GC Wkshps)}, 2021, pp. 1--6.

\bibitem{farsad2018neural}
N.~Farsad and A.~Goldsmith, ``Neural network detection of data sequences in communication systems,'' \emph{IEEE Transactions on Signal Processing}, vol.~66, no.~21, pp. 5663--5678, 2018.

\bibitem{hou2022gru}
J.~Hou, H.~Liu, Y.~Zhang, W.~Wang, and J.~Wang, ``Gru-based deep learning channel estimation scheme for the ieee 802.11 p standard,'' \emph{IEEE Wireless Communications Letters}, 2022.

\bibitem{bai2019deep}
Q.~Bai, J.~Wang, Y.~Zhang, and J.~Song, ``Deep learning-based channel estimation algorithm over time selective fading channels,'' \emph{IEEE Transactions on Cognitive Communications and Networking}, vol.~6, no.~1, pp. 125--134, 2019.

\bibitem{cho2010mimo}
Y.~S. Cho, J.~Kim, W.~Y. Yang, and C.~G. Kang, \emph{MIMO-OFDM wireless communications with MATLAB}.\hskip 1em plus 0.5em minus 0.4em\relax John Wiley \& Sons, 2010.

\bibitem{savaux2017lmmse}
V.~Savaux and Y.~Lou{\"e}t, ``{LMMSE channel estimation in OFDM context: a review},'' \emph{{IET Signal Processing}}, vol.~11, no.~2, pp. 123--134, 2017.

\bibitem{wang2020efficient}
X.~Wang, S.~Ju, X.~Zhang, S.~Ramjee, and A.~El~Gamal, ``Efficient training of deep classifiers for wireless source identification using test {SNR} estimates,'' \emph{IEEE Wireless Communications Letters}, vol.~9, no.~8, pp. 1314--1318, 2020.

\end{thebibliography}

\end{document}